\documentclass[aps,prb,twocolumn,longbibliography] {revtex4-2}
\usepackage{amsmath,graphicx,amsfonts}

\usepackage{times}
\usepackage[colorlinks=true,linkcolor=blue,urlcolor=blue,citecolor=blue, breaklinks=true,hypertexnames=false]{hyperref}

\usepackage{amssymb}
\usepackage{bbm}
\usepackage{bm}

\usepackage[utf8]{inputenc}
\usepackage{comment}
\DeclareUnicodeCharacter{03BC}{\mbox{$\mu$}}

\DeclareUnicodeCharacter{03B4}{$\delta$}
\DeclareUnicodeCharacter{2009}{-}

\usepackage[english]{babel}
\addto\captionsenglish{}
\addto\captionsenglish{}

\makeatletter
\def\bbl@set@language#1{%
	\edef\languagename{%
		\ifnum\escapechar=\expandafter`\string#1\@empty
		\else\string#1\@empty\fi}%
	\@ifundefined{babel@language@alias@\languagename}{}{%
		\edef\languagename{\@nameuse{babel@language@alias@\languagename}}%
	}%
	\select@language{\languagename}%
	\expandafter\ifx\csname date\languagename\endcsname\relax\else
	\if@filesw
	\protected@write\@auxout{}{\string\select@language{\languagename}}%
	\bbl@for\bbl@tempa\BabelContentsFiles{%
		\addtocontents{\bbl@tempa}{\xstring\select@language{\languagename}}}%
	\bbl@usehooks{write}{}%
	\fi
	\fi}
\newcommand{\DeclareLanguageAlias}[2]{%
	\global\@namedef{babel@language@alias@#1}{#2}%
}
\makeatother

\DeclareLanguageAlias{en}{english}

\makeatletter
\def\@bibdataout@aps{%
	\immediate\write\@bibdataout{%
		@CONTROL{%
			apsrev41Control%
			\longbibliography@sw{%
				,author="08",editor="1",pages="1",title="0",year="1"%
			}{%
				,author="08",editor="1",pages="1",title="",year="1"%
			}%
		}%
	}%
	\if@filesw \immediate \write \@auxout {\string \citation {apsrev41Control}}\fi 
}
\makeatother

\def\F {\scriptscriptstyle{F}}
\def\B {\scriptscriptstyle{B}}

\def\BB {\scriptscriptstyle{B\!B}}
\def\FF {\scriptscriptstyle{F\!F}}
\def\BF {\scriptscriptstyle{B\!F}}
\def\FB {\scriptscriptstyle{F\!B}}

\def\L {\scriptscriptstyle{L}}
\def\Y {\scriptscriptstyle{Y}}

\date\today

\begin{document}
\title{Exact solution of the Bose--Fermi mixture along the unnesting line}	\author{Soham Chandak}
\author{Aleksandra Petkovi\'{c}}
\author{Zoran Ristivojevic}
\affiliation{Universit\'{e} de Toulouse, CNRS, Laboratoire de Physique Th\'{e}orique, Toulouse, France}
	
\begin{abstract}
A one-dimensional model of a Bose--Fermi mixture with contact repulsions is  exactly solvable via nested Bethe ansatz. That formal solution is here simplified  along a special line where the Bethe-ansatz equations can be unnested. The model solution is expressed in terms of the known solutions for the Lieb--Liniger model and the Yang--Gaudin model of attractive spin-$\frac{1}{2}$ fermions in the balanced case. We find explicit analytical results for the ground-state energy, the velocities of excitations, and the Drude weights at weak and strong interactions. At weak interactions, the special line corresponds to the parameter values where the isolated systems of free fermions and weakly interacting bosons have equal velocities.
\end{abstract}

\maketitle
	
\section{Introduction}\label{sec1}

Recent advancements in the trapping and cooling of ultracold atoms have enabled the realization of quantum gases in quasi-one-dimensional configurations \cite{catani_2012,spin_charge_guan,schuttelkopf_2026}. Such one-dimensional systems are of particular interest due to the pronounced role of quantum fluctuations \cite{giamarchi}. Although nonperturbative methods such as bosonization successfully describe the low-energy regime, exact theoretical approaches are highly desirable for capturing the physics at arbitrary energy scales. While computing the exact many-body spectrum is generally intractable for most interacting quantum systems, the Bethe-ansatz framework provides exact solutions for several fundamental one-dimensional models. Notable examples include the Heisenberg spin chain \cite{bethe_1931}, the Lieb--Liniger model \cite{lieb_exact_1963a}, the Yang--Gaudin model \cite{gaudin_systeme_1967,yang_exact_1967}, the Hubbard model \cite{lieb_absence_1968}, the Kondo model \cite{kondo_exact}, and the sine-Gordon model \cite{sine_gordon_exact}.

The mere existence of an exact Bethe-ansatz solution does not guarantee straightforward access to physical observables. However, for certain models, remarkable simplifications occur at specific parameter values where the exact solution can be mapped onto simpler theories whose properties are more readily accessible. For instance, along the Luther--Emery line, the sine--Gordon model corresponds to the noninteracting limit of the massive Thirring model \cite{luther_emery_line,coleman_1975}, and thus can be formulated as a free-fermion theory. Analogous mappings allow the single- and multichannel Kondo models to be solved as free-fermion theories along the Toulouse and Emery--Kivelson lines, respectively \cite{toulouse_line,emery_kivelson_line}. Yet another example that can be solved in terms of the free-fermion theory is the limit of zero anisotropy of the Heisenberg model, which is diagonalized using the Jordan--Wigner transformation \cite{lieb_two_1961}.

The one-dimensional Bose--Fermi mixture with contact interactions can be solved exactly using the Bethe ansatz \cite{lai_yang_exact,imambekov_applications_2006}. Building on this exact solution, extensive research has been devoted to understanding the system, including its ground-state properties and low-energy excitation spectrum \cite{batchelor_2005,imambekov_applications_2006,schlottmann_2016,soham_2026}, various correlation functions \cite{imambekov_applications_2006,palacios_2005,patu_2017}, and its thermodynamic behavior \cite{lai_thermodynamics_1974,guan_2012,yin_2009}. However, unlike simpler integrable systems such as the Lieb--Liniger model, the multicomponent nature of the Bose--Fermi mixture necessitates the use of the nested Bethe ansatz. In this framework, the solution introduces auxiliary quasimomenta that do not appear explicitly, for example, in the eigenenergies of the Hamiltonian, but are required to formulate the Bethe equations.

In this article, we identify a specific line in the parameter space where the complicated, nested Bethe-ansatz solution of the Bose--Fermi mixture can be exactly expressed in terms of two simpler, unnested models. Specifically, we demonstrate a decomposition of this system into the Lieb--Liniger model and the balanced Yang--Gaudin model of attractive spin-$\frac{1}{2}$ fermions for the parameters determined by the condition
\begin{align}\label{eq:condition}
n_{\F} \frac{\partial\mu_{\F}}{\partial n_{\F}} = (n_{\B}+n_{\F})\frac{\partial\mu_{\F}}{\partial n_{\B}}.
\end{align}
Here $n_{\F}$ and $n_{\B}$ denote the densities of the fermions and bosons, respectively, in the Bose--Fermi mixture and $\mu_{\F}$ is the chemical potential of fermions. Crucially, this decomposition enables us to make use of well-known relations for the two constituent models to derive explicit results for the Bose--Fermi mixture. Finally, it should be emphasized that unlike the mappings discussed earlier, the decomposition provided here clearly does not correspond to a trivial free-fermion theory, but rather links the system to genuine interacting theories.

The remainder of the paper is organized as follows. In Sec.~\ref{sec:BF_mix_Bethe_sol} we discuss the Bethe-ansatz solution for the Bose--Fermi mixture. Sections~\ref{sec:LL_Bethe_sol} and \ref{sec:YG_Bethe_sol} briefly review the corresponding integral equations for the Lieb--Liniger model and the balanced Yang--Gaudin model of attractive spin-$1/2$ fermions, respectively. In Sec.~\ref{sec:decomp} we establish the exact decomposition of the Bose--Fermi mixture along the special parameter line. Utilizing this mapping, we then derive explicit analytical expressions for the ground-state energy, the excitation velocities, and the Drude weights in both the weak- and strong-coupling regimes. The obtained results are discussed in Sec.~\ref{sec:conclusion}.

\section{Bethe-ansatz solution of the Bose--Fermi mixture}\label{sec:BF_mix_Bethe_sol}
We consider a one-dimensional mixture of bosons and spin-polarized fermions, where the particles interact via contact repulsion. In the case of equal masses of bosons and fermions and equal interparticle interaction strengths, the system admits an exact Bethe-ansatz solution \cite{lai_yang_exact,imambekov_applications_2006}. The corresponding Hamiltonian is given by
\begin{align}
	\label{eq:Hamiltonian}
	H = -\frac{\hbar^2}{2m}\sum_{j=1}^{N}\frac{\partial^2}{\partial x_{j}^2} + \frac{\hbar^2 c}{2m}\sum_{j,l=1\atop j\neq l}^{N}\delta(x_{j}-x_{l}),
\end{align}
where $m$ is the mass of the particles, $N$ is their total number, and $c>0$ describes the repulsion.

The Hamiltonian (\ref{eq:Hamiltonian}) is diagonalized using the Bethe ansatz \cite{lai_yang_exact,imambekov_applications_2006}. For a system with  periodic boundary conditions, the resulting system of equations is given by 
\begin{subequations}\label{eqn:DBAElog}
	\begin{gather}
		Lk_{j} = 2\pi I_{j} + \sum_{l=1}^{N_{\B}}\theta(2k_{j}-2\Lambda_{l}),\quad j = 1,2,\ldots,N,\\
		2\pi J_{l} = \sum_{j=1}^{N}\theta(2\Lambda_{l}-2k_{j}),\quad l = 1,2,\ldots,N_{\B}.
	\end{gather}
\end{subequations}
Here, $N_{\B}$ is the number of bosons, $L$ is the system size, and $k_j$ and $\Lambda_l$ are the quasimomenta and auxiliary quasimomenta, respectively. 
The energy of the Hamiltonian eigenstate is given by $E = (\hbar^2/2m)\sum_{j=1}^{N}k_{j}^2$, while the corresponding momentum is $p = \hbar\sum_{j=1}^{N}k_{j}$. The information about the interactions is encoded in the two-particle scattering phase shift
\begin{align}
	\theta(k) = -2\arctan(k/c).
\end{align}
The numbers $I_{j}$ and $J_{l}$ are integers or odd half-integers, depending on the parity of $N_{\B}$ and $N$. In particular, in the ground state, we have 
\begin{subequations}\label{eqn:quantumnumbers}
	\begin{gather}
		I_{j} = j - \frac{N+1}{2}, \quad j = 1,2,\ldots,N,\\
		J_{l} = l - \frac{N_{\B}+1}{2}, \quad l = 1,2,\ldots,N_{\B}.
	\end{gather}    
\end{subequations}
Other choices for $\{I_{j}\}$ and $\{J_{l}\}$ give rise to excited states.

\subsection{Ground state}

We will study the system in the thermodynamic limit, $L,N,N_{\B}\to \infty$, such that the total density $n=N/L$ and the density of bosons $n_{\B} = N_{\B}/L$ are finite. Consequently, the density of fermions, $n_{\F}=n-n_{\B}$, is also finite. Equations~(\ref{eqn:DBAElog}) for the ground state are then  described by the system of integral equations
\begin{subequations}
	\label{eqn:BetheResc}
	\begin{gather}
		\varrho(x) = \frac{1}{2\pi} + \frac{\lambda}{\pi} \int_{-\eta}^{\eta}dy\,\frac{\varsigma(y)}{\lambda^{2}+(x-y)^2}, \\   
		\varsigma(x) = \frac{\lambda}{\pi}\int_{-1}^{1}dy\,\frac{\varrho(y)}{\lambda^{2}+(x-y)^2}
	\end{gather}
\end{subequations}
that determine the dimensionless densities $\varrho$ of quasimomenta and $\varsigma$ of auxiliary quasimomenta. The dimensionless parameters $\lambda$ and $\eta$ in Eqs.~(\ref{eqn:BetheResc}) are connected to the Fermi (or highest occupied) quasimomentum $Q$ and auxiliary Fermi quasimomentum $B$ via the simple relations
\begin{align}\label{eq:lambdaeta}
	\lambda =\frac{c}{2Q},\quad \eta = \frac{B}{Q}.
\end{align}
Using the result for the particle density $n=Q m_0$, where we have introduced the moments of $\varrho$ as
\begin{align}\label{eq:moments}
m_{j} = \int_{-1}^{1}dx\;\! x^{j}\varrho(x),
\end{align}
we can connect $\lambda$ and $\eta$ with the dimensionless parameters of the model,
\begin{align}\label{eq:parameters}
\gamma=\frac{c}{n},\quad \alpha=\frac{n_{\B}}{n}.
\end{align}
They are
\begin{align}\label{eq:lambda-gamma}
\gamma=\frac{2\lambda}{m_0}
\end{align}
and
\begin{align}\label{eq:eta-alpha}
\alpha=\frac{\int_{-\eta}^{\eta} dx\;\!\varsigma(x)}{m_0}.
\end{align}
While Eqs.~(\ref{eqn:BetheResc}) naturally lead to results expressed in terms of $\lambda$ and $\eta$, in general we prefer to have the same results expressed in terms of $\gamma$ and $\alpha$, which is possible by inverting Eqs.~(\ref{eq:lambda-gamma}) and (\ref{eq:eta-alpha}).

Equations (\ref{eqn:BetheResc}) enable us to find $\varrho(x)$ for $-1\le x\le 1$ and $\varsigma(x)$ for $-\eta\le x\le \eta$. They determine the physical properties of the system. In particular, the ground-state energy $E_0$ can be conveniently expressed as
\begin{align}
	\frac{E_0}{L}=\frac{\hbar^2 n^3}{2m}e(\alpha,\gamma),
\end{align}
where $e(\alpha,\gamma)$ is the dimensionless energy function. It is given in terms of the moments (\ref{eq:moments}) as 
\begin{align}\label{eqn:bf_e_dimless}
	e(\alpha,\gamma) = \frac{m_2}{m_0^3}.
\end{align}
Note that the moments naturally depend on $\lambda$ and $\eta$, which should be eventually expressed in terms of $\gamma$ and $\alpha$ using the relations (\ref{eq:lambda-gamma}) and (\ref{eq:eta-alpha}).

\subsection{Excitation velocities}

The energies of the excited states of the system are also described by the Bethe ansatz. They are encoded in the functions $\Xi(x)$ and $\Omega(x)$ that obey the coupled integral equations
\begin{subequations}\label{eqn:BF_exc_resc}
	\begin{gather}
		\Xi(x) = x + \frac{\lambda}{\pi}\int_{-\eta}^{\eta}dy \frac{\Omega(y)}{\lambda^2 + (x-y)^2}, \\ 
		\Omega(x) = \frac{\lambda}{\pi}\int_{-1}^{1}dy \frac{\Xi(y)}{\lambda^2 + (x-y)^2}. 		
	\end{gather}
\end{subequations}
The momentum $p$ and excitation energy $\varepsilon$ of right-moving elementary excitations obtained by perturbing the quantum numbers $I_{j}$ are given by 
\begin{align}
	p = 2\pi\hbar Q\left|\int_{1}^{\tau}dx\,\varrho(x)\right|, \quad \varepsilon =\frac{\hbar^2 Q^2}{m} \left|\int_{1}^{\tau}dx\,\Xi(x)\right|.
\end{align}
Here $\tau$ is a parameter that controls the momentum and energy of the excitation. For type-I excitations, we have $\tau>1$, and for type-II excitations, $-1<\tau<1$. Similarly, for right-moving elementary excitations obtained by perturbing the quantum numbers $J_{l}$, their momentum and energy are given by
\begin{align}
	p = 2\pi\hbar Q\left|\int_{\eta}^{\tau}dx\,\varsigma(x)\right|, \quad \varepsilon = \frac{\hbar^2Q^2}{m}\left|\int_{\eta}^{\tau}dx\,\Omega(x)\right|.
\end{align}
For type-I excitations, we have $\tau>\eta$ and for type-II excitations, $-\eta<\tau<\eta$. Thus the excitation spectrum of the Bose--Fermi mixture is characterized by four branches. At low momenta, the excitation spectrum is linear and the slope of the branches gives two distinct excitation velocities. They are given by 
\begin{align}\label{eqn:BF_exc_vel_def}
	v_{1} = \frac{\hbar Q}{2\pi m}\frac{\Xi(1)}{\varrho(1)}, \quad v_{2} = \frac{\hbar Q}{2\pi m}\frac{\Omega(\eta)}{\varsigma(\eta)}. 
\end{align}
Like the ground-state energy, the velocities depend on $\alpha$ and $\gamma$.

\section{Bethe-ansatz solution of the Lieb--Liniger model}\label{sec:LL_Bethe_sol}

The Hamiltonian (\ref{eq:Hamiltonian}) also describes identical bosons with contact repulsions, known as the Lieb--Liniger model. In the following, we need some of its properties, which are briefly reviewed here. The Lieb--Liniger model also admits an exact  Bethe-ansatz solution  \cite{lieb_exact_1963a}. For the purpose of notation clarity, we study the particles of mass $m$ in a system of  size $L$ with $N_{\L}$ particles and interaction strength $c_1$, which in the thermodynamic limit is characterized by the density $n_{\L}$.

\subsection{Ground state}

The ground state of the system is described by the Lieb integral equation \cite{lieb_exact_1963a}
\begin{align}\label{eqn:LL_dimensionless}
	\varrho_{\L}(x) = \frac{1}{2\pi} + \frac{\lambda_{1}}{\pi}\int_{-1}^{1}dy\;\!\frac{\varrho_{\L}(y)}{\lambda_{1}^2 + (x-y)^2},
\end{align}
which defines the dimensionless density of quasimomenta $\varrho_{\L}$. In Eq.~(\ref{eqn:LL_dimensionless}), $\lambda_{1}=c_1/Q_{1}$, where $Q_{1}$ is the Fermi quasimomentum. The boson density is $n_{\L}=Q_1 m_{0\L}$, where we have defined the moments
\begin{align}
m_{j_{\L}} = \int_{-1}^{1}dx\;\! x^{j}\varrho_{\L}(x).
\end{align}
Then, the dimensionless Lieb parameter $\gamma_{\L} = c_1/n_{\L}$ is given by
\begin{align}\label{eqn:LL_gamma_def}
	\gamma_{\L} = \frac{\lambda_{1}}{m_{0\L}}.
\end{align}
The ground-state energy $E_{\L}$ can be expressed as 
\begin{align}\label{eqn:LL_e_dimless0}
	\frac{E_{\L}}{L} = \frac{\hbar^2 n_{\L}^3}{2m}e_{\L}(\gamma_{\L}),
\end{align}
where the nontrivial dimensionless function $e_{\L}$ is given by
\begin{align}\label{eqn:LL_e_dimless}
 e_{\L}(\gamma_{\L}) = \frac{m_{2\L}}{(m_{0\L})^3}.
\end{align}
Note that $e_{\L}$ depends only on $\gamma_{\L}$.

\subsection{Excitation velocity}

The excitation spectrum can be expressed via $\,\Xi_{\L}$ which satisfies the integral equation 
\begin{align}
	\Xi_{\L}(x) = x + \frac{\lambda_{1}}{\pi}\int_{-1}^{1}dy\;\!\frac{\Xi_{\L}(y)}{\lambda_{1}^2 + (x-y)^2}.
\end{align}
Then the momentum and energy of the elementary excitations can be, respectively, expressed as 
\begin{align}\label{eq:excLL}
	p_{\L} = 2\pi\hbar Q_1\!\left|\int_{1}^{\tau}dx\;\!\varrho_{\L}(x)\right|, \quad \varepsilon_{\L} =\frac{\hbar^2 Q_1^2}{m} \left|\int_{1}^{\tau}dx\;\!\Xi_{\L}(x)\right|.
\end{align} 
They depend on the parameter $\tau$. For $\tau>1$, Eq.~(\ref{eq:excLL}) describes a right-moving type-I excitation, and for $-1<\tau<1$, it describes a type-II excitation \cite{lieb_exact_1963b}. At low momenta, the excitation spectrum is linear and characterized by a single excitation velocity
\begin{align}\label{eq:25}
	v_{\L} = \frac{\partial \varepsilon_{\L}}{\partial p_{\L}}\Bigg|_{p_{\L}\to 0} = \frac{\hbar Q_1}{2\pi m}\frac{\Xi_{\L}(1)}{\varrho_{\L}(1)}.
\end{align}
Let us note two exact relations involving the excitation velocity. The first one connects the velocity and the density of quasimomenta at the edge \cite{haldane_demo},
\begin{align}\label{eqn:LL_rel_1} 
[2\pi \varrho_{\L}(1)]^2 = \frac{\pi\hbar n_{\L}}{m v_{\L}}.
\end{align}
The second relation involves the velocity and the ground-state energy with its derivatives. It is given by \cite{lieb_exact_1963b}
\begin{subequations}\label{eqn:LL_rel_2}
\begin{align}
	v_{\L}= \frac{\hbar n_{\L}}{m}e_{v\L}(\gamma_{\L}),
\end{align}
where
\begin{align}
e_{v\L}(\gamma_{\L})=\frac{\gamma_{\L}^2}{\sqrt{2}}\left[ \frac{d^2}{d\gamma_{\L}^2} \left(\frac{e_{\L}(\gamma_{\L})}{\gamma_{\L}^2}\right)\right]^{1/2}.
\end{align}
\end{subequations}

\section{Bethe-ansatz solution of the attractive Yang--Gaudin model} \label{sec:YG_Bethe_sol}

Let us now briefly review another exactly solvable model needed in the following sections. It describes a system of spin-$\frac{1}{2}$ fermions with contact attraction, which is known as the Yang--Gaudin model \cite{gaudin_systeme_1967,yang_exact_1967}. It has a Bethe-ansatz solution that has a mathematical structure similar to that of the Lieb--Liniger model. We study the properties of fermions of mass $m$ with the contact attraction of  strength $-c_2$, where we consider $c_2>0$. Focusing on the balanced case with equal numbers of fermions with the two spin projections, and the total number of fermions $N_{\Y}$,  we consider the system of the size $L$ in the thermodynamic limit. It has the density $n_{\Y} = N_{\Y}/L$. 

\subsection{Ground state}

Due to the attraction, fermions of different spin projections are bound into pairs. In the Bethe-ansatz description, this is reflected with the quasimomenta, which occur in complex-conjugate pairs with a constant imaginary part $\pm i c_2/2$. The real part of their dimensionless density, $\varrho_{\Y}$, obeys the equation~\cite{gaudin_systeme_1967,yang_exact_1967}
\begin{align}\label{eqn:YG_rescaled}
	\varrho_{\Y}(x) = \frac{1}{\pi} - \frac{\lambda_{2}}{\pi}\int_{-1}^{1}dy\,\frac{\varrho_{\Y}(y)}{\lambda_{2}^2 + (x-y)^2}.
\end{align}
Here $\lambda_{2} = c_2/Q_{2}$, where $Q_{2}$ is the Fermi quasimomentum. The fermion density is $n_{\Y}=2 Q_2 m_{0\Y} $, where we have introduced the notation
\begin{align}
m_{j\Y} = \int_{-1}^{1}dx\;\! x^{j} \varrho_{\Y}(x).
\end{align}
Then the dimensionless interaction strength $\gamma_{\Y} = c_2/n_{\Y}$ is given by 
\begin{align}\label{eqn:YG_gamma_def}
	\gamma_{\Y} = \frac{\lambda_{2}}{2m_{0\Y}}.
\end{align}
The ground-state energy of the system, $E_{\Y}$,  is given by
\begin{align}\label{eq:binding}
	\frac{E_{\Y}}{L} = \frac{\hbar^2 n_{\Y}^{3}}{2m}\left[-\frac{\gamma_{\Y}^2}{4}+e_{\Y}(\gamma_{\Y})\right].
\end{align}
The first (negative) contribution to the energy describes the binding energy of $N_{\Y}/2$ pairs of fermions. The remaining nontrivial part is described by the dimensionless function $e_{\Y}$ that is given by
\begin{align}\label{eqn:YG_e_dimless}
	e_{\Y}(\gamma_{\Y}) = \frac{m_{2\Y}}{4(m_{0\Y})^3},
\end{align}
which depends only on $\gamma_{\Y}$.

\subsection{Excitation velocity}

Information about the excitation spectrum is contained in $\Xi_{\Y}$ that obeys  the equation 
\begin{align}
\Xi_{\Y}(x) =x -\frac{\lambda_{2}}{\pi} \int_{-1}^{1}dy\;\!\frac{\Xi_{\Y}(y)}{\lambda_{2}^2 + (x-y)^2}.
\end{align}
The momentum $p_{\Y}$ and energy $\varepsilon_{\Y}$ of elementary excitations can be expressed as 
\begin{align}\label{eq:pe-YG}
	p_{\Y} = 2\pi\hbar Q_2\!\left|\int_{1}^{\tau}dx\;\!\varrho_{\Y}(x)\right|, \quad \varepsilon_{\Y} = \frac{2\hbar^2 Q_2^2}{m}\left|\int_{1}^{\tau}dx\;\!\Xi_{\Y}(x)\right|.
\end{align} 
Here, $\tau$ is a parameter that determines $p_{\Y}$ and $\varepsilon_{\Y}$. For $\tau>1$, there are type-I excitations and for $-1<\tau<1$, there are type-II excitations. Note the difference of a multiplicative factor of $2$ when comparing Eqs.~(\ref{eq:excLL}) and (\ref{eq:pe-YG}). It arises because, for attractive fermions in the Yang--Gaudin model, the fermions are bound into pairs and each of their constituents contributes to the excitation energy. On the other hand, $\varrho_{\Y}$ already describes pairs as its integral gives the pair density and thus there is no extra factor of $2$ in the momentum in Eq.~(\ref{eq:pe-YG}). At low momenta, the excitation spectrum is linear and characterized by the excitation velocity $v_{\Y}$. It is given by
\begin{align}\label{eq:35}
	v_{\Y} = \frac{\partial \varepsilon_{\Y}}{\partial p_{\Y}}\Bigg|_{p_{\Y}\to 0} = \frac{\hbar Q_2}{\pi m}\frac{\Xi_{\Y}(1)}{\varrho_{\Y}(1)}.
\end{align}
Similar to the Lieb--Liniger model, we note the following exact relations for the excitation velocity:
\begin{align}\label{eqn:YG_rel_1}
[2\pi \varrho_{\Y}(1)]^2 = \frac{\pi \hbar n_{\Y}}{m v_{\Y}},
\end{align}
and
\begin{subequations}\label{eqn:YG_rel_2}
\begin{align}
	v_{\Y} &= \frac{\hbar n_{\Y}}{m}e_{v\Y}(\gamma_{\Y}),
\end{align}
where
\begin{align}
e_{v\Y}(\gamma_{\Y})=\frac{\gamma_{\Y}^2}{\sqrt{2}}\left[ \frac{d^2}{d\gamma_{\Y}^2} \left(\frac{e_{\Y}(\gamma_{\Y})}{\gamma_{\Y}^2}\right)\right]^{1/2}.
\end{align}
\end{subequations}
These relations can be derived analogously to the ones in the Lieb--Liniger model. The velocity $v_{\Y}$ describes the linear low-energy dispersion of bound pairs, corresponding to the density excitations \cite{fuchs_exactly_2004}.

\section{The solution of the nested system at $\eta=1$}\label{sec:decomp}

The main goal of this work is to provide an explicit exact solution for the Bose--Fermi mixture along a special line defined by Eq.~(\ref{eq:condition}). As will be shown in Sec.~\ref{sec54}, it corresponds to the special choice $\eta=1$ in the system of integral equations~(\ref{eqn:BetheResc}). In that case, the two Fermi rapidities are equal, $Q=B$; see Eq.~(\ref{eq:lambdaeta}).

\subsection{Ground state}\label{sec5}

The system (\ref{eqn:BetheResc}) at $\eta=1$ can be decoupled by introducing the functions $\varrho_{\pm}(x) = \varrho(x) \pm \varsigma(x)$. Indeed, by adding and subtracting Eqs.~(\ref{eqn:BetheResc}), we obtain the pair of equations
\begin{align}\label{eqn:add_sub}
	\varrho_{\pm}(x) = \frac{1}{2\pi} \pm \frac{\lambda}{\pi}\int_{-1}^{1}dy\, \frac{\varrho_{\pm}(y)}{\lambda^2 + (x-y)^2}.
\end{align}
Up to a trivial multiplication, they are identical to  Eqs.~(\ref{eqn:LL_dimensionless}) and (\ref{eqn:YG_rescaled}) that refer to the ground state of the Lieb--Liniger model and the balanced case of the attractive  Yang--Gaudin model, provided 
\begin{align}\label{eq:lambdasequal}
\lambda_{1} = \lambda_{2} = \lambda
\end{align}
 is used. In particular, we can identify 
\begin{align}
	\varrho_{+}(x) = \varrho_{\L}(x), \quad \varrho_{-}(x) = \frac{\varrho_{\Y}(x)}{2},
\end{align}
enabling us to find  
\begin{align}\label{eqn:eta1_gs_rel}
	\varrho(x) = \frac{\varrho_{\L}(x)}{2} + \frac{\varrho_{\Y}(x)}{4}, \quad \varsigma(x) = \frac{\varrho_{\L}(x)}{2} - \frac{\varrho_{\Y}(x)}{4}.
\end{align}
Therefore, at $\eta=1$, the nested Bethe ansatz for the Bose--Fermi mixture greatly simplifies. It reduces to two simpler Bethe-ansatz solutions for the Lieb--Liniger and fully paired Yang-Gaudin models.

\begin{figure}
\includegraphics[width=\columnwidth]{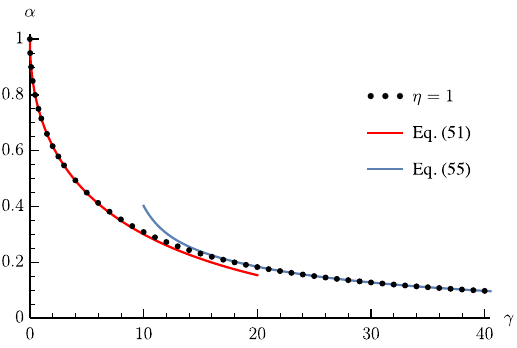}
\caption{The dots represent the numerical solution of the curve $\eta = 1$ in the $\alpha$-$\gamma$ plane. The solid curves are its analytical forms (\ref{eqn:gamma_wk_fn_alpha}) and  (\ref{eqn:gamma_str_fn_alpha}) in the two limiting cases of small and large $\gamma$.}
\label{fig:eta1}
\end{figure}

From the definitions (\ref{eq:lambda-gamma}), (\ref{eq:eta-alpha}), (\ref{eqn:LL_gamma_def}), and (\ref{eqn:YG_gamma_def}), we obtain
\begin{align}\label{eqn:eta1_dim_less_parameters_rel}
	\gamma_{\L} = \frac{\gamma}{2(1+\alpha)}, \quad \gamma_{\Y} = \frac{\gamma}{8(1-\alpha)},
\end{align} 
connecting the dimensionless parameters. Note that along the line $\eta=1$, there is only one free parameter, for example $\gamma$, which fixes the values of $\alpha$, $\gamma_{\L}$, and $\gamma_{\Y}$. The values of $\gamma$ and $\alpha$ are determined by Eqs.~(\ref{eq:lambda-gamma}) and (\ref{eq:eta-alpha}). The corresponding curve is shown in Fig.~\ref{fig:eta1}. It shows that as $\alpha$ approaches $1$, which is the limit where the fermion density vanishes, the interaction strength is small, $\gamma\ll 1$. In the opposite limit of small $\alpha$, which corresponds to a small boson density, we see that the constraint $\eta=1$ corresponds to the strong interactions in the system,  $\gamma\gg 1$.

Using the definitions (\ref{eqn:bf_e_dimless}), (\ref{eqn:LL_e_dimless}), and (\ref{eqn:YG_e_dimless}) of the ground-state energies in the respective models, we find the relation
\begin{align}\label{eqn:eta1_energy_relation}
	e(\alpha,\gamma) ={}& \frac{(1+\alpha)^3}{2}e_{\L}\pmb{\biggl(}\frac{\gamma}{2(1+\alpha)}\pmb{\biggr)} \notag \\ &+ 8(1-\alpha)^3 e_{\Y}\pmb{\biggl(}\frac{\gamma}{8(1-\alpha)}\pmb{\biggr)}.
\end{align}
Therefore, the ground-state energy of the Bose--Fermi mixture is a weighted sum of the energies of the Lieb--Liniger and Yang--Gaudin models, where the latter are realized at special interaction strengths. Note that the binding energy term of Eq.~(\ref{eq:binding}) does not participate in Eq.~(\ref{eqn:eta1_energy_relation}).

Let us analytically find the relation between $\alpha$ and $\gamma$ along the line $\eta=1$. In order to achieve this, we use the relation between the Fermi quasimomentum and the ground-state energy function. For the Lieb--Liniger model, it can be expressed as \cite{ristivojevic_method_2022,zoran_lieb_liniger_review}
\begin{align}\label{eqn:g_rel_eta_1}
	\lambda_1 = \frac{\sqrt{\gamma_{\L}}}{2g_{1}(\gamma_{\L})},
\end{align}
where $g_{1}(\gamma_{\L})$ is a solution of the differential equation
\begin{align}
	\frac{d}{d\gamma_{\L}}\left(\frac{g_{1}(\gamma_{\L})}{\sqrt{\gamma_{\L}}}\right) = -\sqrt{\frac{1}{8}\frac{d^2}{d\gamma_{\L}^2}\left(\frac{e_{\L}(\gamma_{\L})}{\gamma_{\L}^2}\right)}.
\end{align}
An analogous relation for the Yang--Gaudin model is given by
\begin{align}\label{eqn:g_rel_eta_2}
	\lambda_2 = \frac{\sqrt{\gamma_{\Y}}}{2g_{2}(\gamma_{\Y})},
\end{align}
where $g_{2}(\gamma_{\Y})$ is a solution of the differential equation
\begin{align}
	\frac{d}{d\gamma_{\Y}}\left(\frac{g_{2}(\gamma_{\Y})}{\sqrt{\gamma_{\Y}}}\right) = -\sqrt{\frac{1}{8}\frac{d^2}{d\gamma_{\Y}^2}\left(\frac{e_{\Y}(\gamma_{\Y})}{\gamma_{\Y}^2}\right)}.
\end{align}
Since there are no known exact expressions for $e_{\L}(\gamma_{\L})$ and $e_{\Y}(\gamma_{\Y})$ in terms of elementary functions, we can use their series expansions in the regimes of weak and strong interactions \cite{marino_resurgence_2019,zoran_lieb_liniger_review}. Apart from that, the linear differential equations for $g_1(\gamma_{\L})$ and $g_2(\gamma_{\Y})$ require one integration constant. At strong interactions, it is zero due to the physical requirement of finite Fermi quasimomenta. At weak interactions, the determination of the integration constants is significantly more complicated. For the Lieb--Liniger model, it is done in Ref.~\cite{zoran_lieb_liniger_review}, while in the Yang--Gaudin model, it follows from the zeroth moment of the solution of Eq.~(\ref{eqn:YG_rescaled}) that is calculated perturbatively in Refs.~\cite{tracy_ground_2018,marino_resurgence_2019}.

At weak interactions, $\gamma \ll 1$, we have $\gamma_{\L}\ll 1$ and $\gamma_{\Y} \ll 1$ as will be \textit{a posteriori} verified. The series expansion of the function $g_{1}$ is given by \cite{zoran_lieb_liniger_review} 
\begin{align}
	g_{1}(\gamma_{\L}) ={}& 1 - \frac{\sqrt{\gamma_{\L}}}{4\pi}\left[\ln\left(\frac{32\pi}{\sqrt{\gamma_{\L}}}\right)-1\right]+\frac{\gamma_{\L}}{32\pi^2}+O(\gamma_{\L}^{3/2}).
\end{align}
For the other function, we obtain
\begin{align}
	g_{2}(\gamma_{\Y}) ={}&  \frac{\pi}{4\sqrt{\gamma_{\Y}}} - \frac{\sqrt{\gamma_{\Y}}}{4\pi}\left[\ln\left(\frac{\pi^2}{2\gamma_{\Y}}\right)+1\right] + \frac{\gamma_{\Y}^{3/2}}{8\pi^3} \notag \\ &+ O(\gamma_{\Y}^{5/2}).
\end{align}
Assuming a power-law series, from Eqs.~(\ref{eq:lambdasequal}), (\ref{eqn:g_rel_eta_1}), and (\ref{eqn:g_rel_eta_2}), we now obtain
\begin{align}\label{eqn:gLL_gYG_rel}
\gamma_{\L} ={}& \frac{16}{\pi^2}\gamma_{\Y}^2 - \frac{64[2\ln(2)-1]}{\pi^4}\gamma_{\Y}^3 \notag \\ &+ \frac{128\{6[\ln(2)]^2-7\ln(2)+2\}}{\pi^6}\gamma_{\Y}^4+O(\gamma_{\Y}^5).
\end{align}
Equations~(\ref{eqn:gLL_gYG_rel}) and (\ref{eqn:eta1_dim_less_parameters_rel}) now confirm that the smallness of $\gamma$ implies smallness of $\gamma_{\L}$ and $\gamma_{\Y}$. Eventually, using Eq.~(\ref{eqn:gLL_gYG_rel}) and Eq.~(\ref{eqn:eta1_dim_less_parameters_rel}), we obtain
\begin{align}\label{eqn:gamma_wk_fn_alpha}
	\alpha = 1-\frac{\sqrt{\gamma}}{\pi}+\frac{\ln(2)}{2\pi^2}\gamma -\frac{\ln(2)}{16\pi^3}\gamma^{3/2} + O(\gamma^2).
\end{align}
This is the analytical form of the curve $\eta=1$ in the vicinity of $\alpha=1$; see Fig.~\ref{fig:eta1}. We can then use the obtained $\alpha(\gamma)$ dependence of Eq.~(\ref{eqn:gamma_wk_fn_alpha}) and find the energy (\ref{eqn:eta1_energy_relation}). It is given by
\begin{align}\label{eqn:eta1_en_wk_alpha}
e^*(\gamma)={}& \gamma - \frac{\gamma^{3/2}}{\pi} +\frac{\pi^2-12\ln(2)}{24\pi^2}\gamma^2 \notag \\ &+ \frac{32 [\ln(2)]^2 + 4 \ln(2)+3\zeta{(3)}}{64\pi^3}\gamma^{5/2} + O(\gamma^3),
\end{align}
where we have introduced the abbreviation $e^*(\gamma)\equiv	e(\alpha,\gamma)$ with $\alpha$ taken along the $\eta=1$ line, and $\zeta$ denotes the Riemann zeta function. Because $\alpha$ is in the vicinity of $1$ for $\gamma\ll 1$, the system has a very small density of fermions. In that case the leading-order term in Eq.~(\ref{eqn:eta1_en_wk_alpha}) coincides with the result for the Lieb--Liniger model \cite{lieb_exact_1963a}, as expected.

\begin{figure}
\includegraphics[width=\columnwidth]{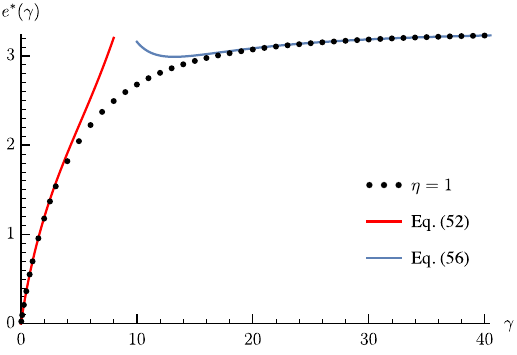}
\caption{The dots represent numerically exact results for the energy function $e^*(\gamma)\equiv e(\alpha,\gamma)$ along the line $\eta=1$ obtained by solving the integral equations. The solid curves are the two analytical expressions (\ref{eqn:eta1_en_wk_alpha}) and (\ref{eqn:eta1_en_str_alpha}) for the limiting cases of weak and strong interactions.}
\label{fig:eta1_en}
\end{figure}

In the regime of strong interactions, $\gamma \gg 1$, we have $\gamma_{\L} \gg 1$ and $\gamma_{\Y} \gg 1$. The function $g_1$ is given by \cite{zoran_lieb_liniger_review}
\begin{align}\label{eqn:g1_str}
	g_{1}(\gamma_{\L}) = \frac{\pi}{2\sqrt{\gamma_{\L}}}\left[1-\frac{2}{\gamma_{\L}}+\frac{4}{\gamma_{\L}^2}+\frac{4\pi^2-24}{3\gamma_{\L}^3}+O(\gamma_{\L}^{-4})\right].
\end{align} 
For the other function, we find
\begin{align}\label{eqn:g2_str}
	g_{2}(\gamma_{\Y}) = \frac{\pi}{8\sqrt{\gamma_{\Y}}}\left[1+\frac{1}{2\gamma_{\Y}}+\frac{1}{4\gamma_{\Y}^2}-\frac{\pi^2-6}{48\gamma_{\Y}^3}+O(\gamma_{\Y}^{-4})\right].
\end{align}
Using Eqs.~(\ref{eqn:g_rel_eta_1}), (\ref{eqn:g_rel_eta_2}), and (\ref{eqn:eta1_dim_less_parameters_rel}), we obtain
\begin{align}\label{eqn:gamma_str_fn_alpha}
	\alpha ={}& \frac{4}{\gamma}-\frac{32(\pi^2+6)}{3\gamma^3} + \frac{512(2\pi^4+25\pi^2+60)}{15\gamma^5}+O(\gamma^{-6}).
\end{align}
This is the analytical form of the $\eta=1$ line at $\gamma\gg 1$; see Fig.~\ref{fig:eta1}. Using Eq.~(\ref{eqn:gamma_str_fn_alpha}), the energy (\ref{eqn:eta1_energy_relation}) becomes 
\begin{align}\label{eqn:eta1_en_str_alpha}
e^*(\gamma) = \frac{\pi^2}{3}\left(1-\frac{32}{\gamma^2}+\frac{256(3\pi^2+25)}{5\gamma^4}\right)+O(\gamma^{-5}).
\end{align}
As expected, the result (\ref{eqn:eta1_en_str_alpha}) at leading order coincides with that of free fermions, since in this case $\alpha$ is very small, which denotes a small density of bosons. The system predominantly consists of fermions, for which the contact interaction is not effective due to the Pauli principle and thus they behave as noninteracting particles. The energy $e^*(\gamma)$ is plotted in Fig.~\ref{fig:eta1_en}. It increases monotonically from $0$ at $\gamma = 0$ to the maximal value of $\pi^2/3$ at $\gamma \to \infty$ as one follows the $\eta=1$ line.

\subsection{Excitation spectrum}
When $\eta=1$ in Eq.~(\ref{eqn:BF_exc_resc}), we can express $\Xi$ and $\Omega$ in terms of $\Xi_{\L}$ and $\Xi_{\Y}$, which govern the excitation spectrum of the Lieb--Liniger and the balanced attractive Yang--Gaudin models. We find
\begin{align}\label{eqn:eta1_ex_rel}
	\Xi(x) = \frac{\Xi_{\L}(x)+\Xi_{\Y}(x)}{2}, \quad \Omega(x) = \frac{\Xi_{\L}(x)-\Xi_{\Y}(x)}{2}.
\end{align}
This enables us to express the elementary excitations in the Bose--Fermi mixture as a combination of elementary excitations of the two simpler models.

From the definition (\ref{eqn:BF_exc_vel_def}) of the excitation velocities for the Bose--Fermi mixture and the relations (\ref{eq:25}), (\ref{eqn:LL_rel_1}), (\ref{eq:35}), (\ref{eqn:YG_rel_1}), (\ref{eqn:eta1_gs_rel}), and (\ref{eqn:eta1_ex_rel}), we can express the excitation velocities for the Bose--Fermi mixture as 
\begin{align}\label{eqn:eta1_v1_v2_rel}
	v_{1,2}={}&\frac{\frac{2Q}{Q_1}v_{\L}\varrho_{\L}(1)\pm \frac{Q}{Q_2}v_{\Y}\varrho_{\Y}(1)}{2\varrho_{\L}(1)\pm \varrho_{\Y}(1)}.
\end{align}
Then, using the relations (\ref{eqn:LL_rel_2}) and (\ref{eqn:YG_rel_2}) for the excitation velocities in the Lieb--Liniger and Yang--Gaudin models, and the definitions $\lambda=\gamma n/2Q$, $\lambda_{1}=\gamma_{\L}n_{\L}/Q_1$, and $\lambda_{2}=\gamma_{\Y}n_{\Y}/Q_2$, we can express Eq.~(\ref{eqn:eta1_v1_v2_rel}) as 
\begin{align}
	v_{1,2} = \frac{\hbar n}{m} \frac{\sqrt{e_{v\L}e_{v\Y}}}{2\sqrt{e_{v\Y}}\pm\sqrt{e_{v\L}}} \left(\frac{\gamma\lambda_{1}}{\gamma_{\L}\lambda}\sqrt{e_{v\L}}\pm \frac{\gamma\lambda_2}{2\gamma_{\Y} \lambda}\sqrt{e_{v\Y}}\right). 
\end{align}
Eventually, using the connections (\ref{eq:lambdasequal}) and (\ref{eqn:eta1_dim_less_parameters_rel}), we obtain
\begin{align}\label{eqn:eta1_v1_v2_vLL_vYG}
	v_{1,2} = \frac{\hbar n}{m} \frac{2(1+\alpha)\sqrt{e_{v\L}}\pm 4(1-\alpha)\sqrt{e_{v\Y}}}{2\sqrt{e_{v\Y}}\pm\sqrt{e_{v\L}}} \sqrt{e_{v\L}e_{v\Y}}. 
\end{align}
Here $e_{v\L}\equiv e_{v\L}(\gamma_{\L})$ and $e_{v\Y}\equiv e_{v\Y}(\gamma_{\Y})$ as defined in Eqs.~(\ref{eqn:LL_rel_2}) and (\ref{eqn:YG_rel_2}), respectively.

In the weakly interacting limit, $\gamma\ll 1$, for the excitation velocities in the Bose--Fermi mixture we obtain
\begin{align}\label{eqn:eta1_v_wk}
	v_{1,2} ={}& \frac{\hbar n}{m}\sqrt{\gamma}\biggl(1 \pm \frac{\gamma^{1/4}}{2\pi^{1/2}}-\frac{5\gamma^{1/2}}{8\pi}\mp\frac{2\ln(2)-1}{4\pi^{3/2}}\gamma^{3/4}\notag \\ &+\frac{48\ln(2)-25}{128\pi^2}\gamma \mp \frac{13\ln(2)-8}{64\pi^{5/2}}\gamma^{5/4} + O(\gamma^{3/2}) \biggr).
\end{align}
Note that Eq.~(\ref{eqn:eta1_v_wk}) describes two velocities. 
At leading order, $O(\sqrt\gamma)$, the Lieb--Liniger sound velocity and the free-fermion Fermi velocity [corresponding to the density $n_{\F}=(1-\alpha) n=\sqrt\gamma n/\pi$] are equal, so the two modes are degenerate. The subleading-order terms lift this degeneracy.

In the strongly interacting limit, $\gamma\gg 1$, we find 
\begin{subequations}\label{eqn:eta1_v_str}
\begin{align}
	v_{1} &= \frac{\pi\hbar n}{m}\left(1-\frac{32}{\gamma^2}+\frac{640(\pi^2+6)}{3\gamma^4}\right)+O(\gamma^{-5}), \\ 
	v_{2} &= \frac{\pi\hbar n}{m}\frac{8\pi^2}{3\gamma^2}\left(1-\frac{16(7\pi^2+60)}{15\gamma^2}\right)+O(\gamma^{-5}).
\end{align}	
\end{subequations}
At leading order, $v_1$ corresponds to the free fermion velocity. This is expected as at $\gamma\gg 1$, we have $\alpha\to 0$ and thus negligible boson density in such a system. The presence of one velocity in such a system is also reflected in $v_2$, which vanishes as $\gamma\to\infty$.

\begin{figure}
\includegraphics[width=\columnwidth]{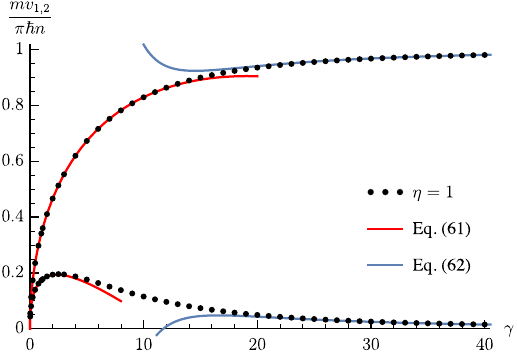}
\caption{Plot of the velocities of excitations along the line defined by $\eta=1$. The dots represent numerically exact results obtained by solving the integral equations. The solid curves, which represent the corresponding analytical forms in the regimes of weak and strong interactions, are given by Eqs.~(\ref{eqn:eta1_v_wk}) and (\ref{eqn:eta1_v_str}).}
\label{fig:eta1_v1}
\end{figure}

Figure~\ref{fig:eta1_v1} shows the excitation velocities $v_{1}$ and $v_{2}$ as a function of $\gamma$ along the line $\eta=1$. We see that the excitation velocity $v_{1}$ monotonically increases as a function of $\gamma$ and approaches the value $\pi\hbar n/m$ as $\gamma \to \infty$. The excitation velocity $v_{2}$, on the other hand, is nonmonotonic as a function of $\gamma$, attaining a maximal value at some intermediate value of $\gamma$.  The velocity $v_2$ approaches $0$ in both limits $\gamma\to 0$ and $\gamma \to \infty$, reflecting that the mixture becomes a single-component system.

\subsection{Drude weights}

Knowledge of the excitation velocities and the ground-state energy enables us to directly find the Drude weights. The Drude weight quantifies  the ability of a system to sustain a particle current without dissipation. Following Kohn \cite{kohn_drude_1964}, the Drude weight can be calculated by applying twisted boundary conditions to the system. For a Bose--Fermi mixture we thus have
\begin{equation}
	\begin{aligned}
		\Psi(\ldots,x_{j}+L,\ldots) ={}& \Psi(\ldots,x_{j},\ldots)\\ &\times\begin{cases}
			e^{i\varphi_{\B}} & 1\leq j\leq N_{\B},\\
			e^{i\varphi_{\F}} & N_{\B}+1\leq j\leq N.
		\end{cases}
	\end{aligned}    
\end{equation}
Then the Drude weight matrix
\begin{align}
	\mathcal{D} = \begin{pmatrix}
		\mathcal{D}_{\FF} & \mathcal{D}_{\FB} \\
		\mathcal{D}_{\BF} & \mathcal{D}_{\BB} \\
	\end{pmatrix}
\end{align}
is defined via 
\begin{align}
	\mathcal{D}_{ss'} = \frac{\pi L}{\hbar}\frac{\partial^2 E_{t}}{\partial \varphi_{s}\partial\varphi_{s'}}\Bigg|_{\varphi_{s},\varphi_{s'}\to0}, \quad s,s' \in \{B,F\},
\end{align}
where $E_{t}$ is the ground-state energy of the system with twisted boundary conditions. The Drude weight matrix is symmetric.

In Ref.~\cite{soham_2026}, we derived that the Drude weight matrix obeys the exact relation 
\begin{align}\label{eqn:DW_matrix_exact}
	\mathcal{D} = \mathcal{Z}\mathcal{V}\mathcal{Z}^{T}.
\end{align}
Here, we have defined 
\begin{align}
	\mathcal{V} = \begin{pmatrix}
		v_{1} & 0 \\
		0 & v_{2} \\
	\end{pmatrix}, \quad \mathcal{Z} = 2\pi\begin{pmatrix}
	\varrho(1) - \tilde{\varsigma}(1) & \varsigma(\eta)-\tilde{\varrho}(\eta)\\
	\tilde{\varsigma}(1) & \tilde{\varrho}(\eta) \\
	\end{pmatrix},
\end{align}
where $\tilde{\varrho}$ and $\tilde{\varsigma}$ are the solutions of the coupled integral equations
\begin{subequations}\label{eq:dualeqs}
	\begin{gather}
		\tilde{\varrho}(x) = \frac{1}{2\pi} + \frac{\lambda}{\pi} \int_{-1}^{1}dy\,\frac{\tilde{\varsigma}(y)}{\lambda^{2}+(x-y)^2}, \\   
		\tilde{\varsigma}(x) = \frac{\lambda}{\pi}\int_{-\eta}^{\eta}dy\,\frac{\tilde{\varrho}(y)}{\lambda^{2}+(x-y)^2}.
	\end{gather}    
\end{subequations}
Due to Galilean invariance, the Drude weight matrix elements obey \cite{soham_2026}
\begin{subequations}\label{eq:sumrulesDrude}
\begin{gather}
\mathcal{D}_{\FF} +\mathcal{D}_{\FB}=\frac{\pi\hbar n_{\F}}{m},\\
\mathcal{D}_{\BB} +\mathcal{D}_{\FB}=\frac{\pi\hbar n_{\B}}{m}.
\end{gather}
\end{subequations}
Thus, the Drude weight matrix has only one independent matrix element, while the others follow from the sum rules (\ref{eq:sumrulesDrude}). It is therefore sufficient to consider $\mathcal{D}_{\FB}$, for example.

For our case of $\eta=1$, the integration limits in Eqs.~(\ref{eqn:BetheResc}) and (\ref{eq:dualeqs}) coincide and thus $\varrho = \tilde{\varrho}$ and $\varsigma = \tilde{\varsigma}$. This simplification used in Eq.~(\ref{eqn:DW_matrix_exact}) enables us to find 
\begin{align}\label{eq:Dbfgen}
	\mathcal{D}_{\FB} = 4\pi^2[\varrho(1)-\varsigma(1)][v_1\varsigma(1)-v_2\rho(1)].
\end{align}
Using Eq.~(\ref{eqn:eta1_v1_v2_vLL_vYG}), we eventually obtain
\begin{align}
	\mathcal{D}_{\FB} &= \frac{\pi\hbar n}{m}\frac{4(1-\alpha)e_{v\Y}-(1+3\alpha)e_{v\L}}{4e_{v\Y}-e_{v\L}},
\end{align}
where $e_{v\L}\equiv e_{v\L}(\gamma_{\L})$ and $e_{v\Y}\equiv e_{v\Y}(\gamma_{\Y})$ as defined in Eqs.~(\ref{eqn:LL_rel_2}) and (\ref{eqn:YG_rel_2}), respectively. 

\begin{figure}
\includegraphics[width=\linewidth]{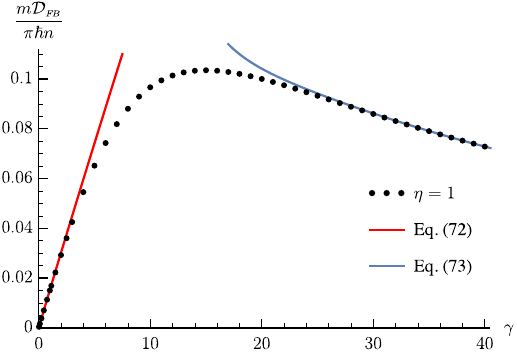}
\caption{The dots are numerically exact values of $\mathcal{D}_{\FB}$ obtained by studying the corresponding integral equations at $\eta = 1$. The curves represent the analytical expansions  (\ref{eqn:eta1_dff_wk}) and (\ref{eqn:eta1_dff_str}) in the limiting cases.}
\label{fig:eta1_Dbf}
\end{figure}

In the weakly interacting regime, the Drude weight $\mathcal{D}_{\FB}$ takes the explicit form
\begin{align}
\label{eqn:eta1_dff_wk}
	\mathcal{D}_{\FB} ={}& \frac{\pi\hbar n}{m}\left[\frac{1-\ln(2)}{2\pi^2}\gamma -\frac{3\ln(2)-2}{8\pi^3}\gamma^{3/2} +O(\gamma^{2})\right].
\end{align}
Interestingly, unlike the velocities (\ref{eqn:eta1_v_wk}), the Drude weight matrix elements are power series in $\gamma^{1/2}$. This occurs because odd powers of $\gamma^{1/4}$ disappear in the product $v_1 v_2$, which enters the Drude weight matrix elements as a multiplicative factor. In the opposite regime of strong interactions, we obtain
\begin{align}
\label{eqn:eta1_dff_str}
	\mathcal{D}_{\FB} ={}& \frac{\pi\hbar n}{m}\Bigg(\frac{4}{\gamma}- \frac{8(\pi^2+6)}{3\gamma^2}-\frac{32(\pi^2+6)}{3\gamma^3}\notag \\ &+\frac{128(7\pi^4+90\pi^2+180)}{45\gamma^4}\Bigg)+ O(\gamma^{-5}). 
\end{align}
The remaining Drude weights $\mathcal{D}_{\FF}$ and $\mathcal{D}_{\BB}$ follow from the sum rules (\ref{eq:sumrulesDrude}). The explicit results obtained here at strong interactions along the line $\eta=1$, for the energy (\ref{eqn:eta1_en_str_alpha}), the velocities (\ref{eqn:eta1_v_str}), and the Drude weight (\ref{eqn:eta1_dff_str}), can also be obtained from the results calculated at strong interactions and arbitrary $\alpha$ \cite{chandak_unpub} by substituting there the relation (\ref{eqn:gamma_str_fn_alpha}). In Fig.~\ref{fig:eta1_Dbf} we have plotted the Drude weight $\mathcal{D}_{\FB}$ along the line $\eta=1$ as a function of $\gamma$. We see that the Drude weight has a nonmonotonic behavior, vanishing in both limiting cases of weak and strong interactions. Note that the terms proportional to $\gamma^{-3}$ and $\gamma^{-4}$ in Eq.~(\ref{eqn:eta1_dff_str}) are important to be kept together from the practical point of view as they have comparable numerical values, significantly improving the agreement between the exact result and the analytical expression (\ref{eqn:eta1_dff_str}) for the range of values of $\gamma$ displayed in Fig.~\ref{fig:eta1_Dbf}.

\subsection{Equivalent form of the condition $\eta=1$}\label{sec54}

The ground-state energy $E_0$ enables us to study the compressibility matrix,
\begin{align}
	\mathcal{M} = \begin{pmatrix}
		\mathcal{M}_{\FF} & \mathcal{M}_{\FB} \\
		\mathcal{M}_{\BF} & \mathcal{M}_{\BB} \\
	\end{pmatrix},
\end{align} 
where the matrix elements are conveniently defined by
\begin{align}
\mathcal{M}_{ss'} =\frac{1}{\pi\hbar} \frac{\partial \mu_{s}}{\partial n_{s'}}, \quad s,s' \in \{B,F\}.
\end{align}
Here, $\mu_{\F}$ and $\mu_{\B}$ are chemical potentials defined via
\begin{align}
	\mu_{\F} = \frac{\partial E_0}{\partial N_{\F}}, \quad \mu_{\B} = \frac{\partial E_0}{\partial N_{\B}},
\end{align}
where $N_{\F}$ is the fermion number. The matrix $\mathcal{M}$ is symmetric. In Ref.~\cite{soham_2026} we derived that the compressibility matrix satisfies the exact relation
\begin{align}\label{eqn:DW_matrix_exact1}
\mathcal{M} = (\mathcal{Z}^{-1})^{T}V\mathcal{Z}^{-1}.
\end{align}
Inverting Eq.~(\ref{eqn:DW_matrix_exact1}) and using Eqs.~(\ref{eqn:DW_matrix_exact}) and (\ref{eq:sumrulesDrude}), for $\eta=1$ we obtain the two constraints
\begin{subequations}\label{eq:1}
\begin{gather}
	\frac{\pi\hbar n}{m} = \frac{v_{1}v_{2}}{\det{\mathcal{M}}}\mathcal{M}_{\FF}, \\
	\frac{\pi\hbar n_{\B}}{m} = \frac{v_{1}v_{2}}{\det{\mathcal{M}}}(\mathcal{M}_{\FF}-\mathcal{M}_{\BF}).
\end{gather}
\end{subequations}
Since $\det{\mathcal{M}}$ and the product $v_1v_2$ are nonzero at any finite interaction \cite{soham_2026}, dividing the two expressions (\ref{eq:1}) we obtain Eq.~(\ref{eq:condition}). Therefore, the line $\eta=1$, which denotes equal Fermi quasimomentum of the auxiliary quasimomenta and the quasimomenta, can be expressed in the equivalent form (\ref{eq:condition}), containing the thermodynamic quantities of the system.

\section{Discussion}\label{sec:conclusion}

Here we give some additional comments. In combination with Eq.~(\ref{eqn:DW_matrix_exact}), the sum rules (\ref{eq:sumrulesDrude}) enable us to express the velocities of excitations as
\begin{subequations}
\begin{gather}
v_1=\frac{\hbar n}{4\pi m}\frac{\varrho(1)-\alpha\varsigma(1)}{\varrho(1)\left\{[\varrho(1)]^2-[\varsigma(1)]^2\right\}},\\
v_2=\frac{\hbar n}{4\pi m}\frac{\alpha\varrho(1)-\varsigma(1)}{\varsigma(1)\left\{[\varrho(1)]^2-[\varsigma(1)]^2\right\}}.
\end{gather}
\end{subequations}
Therefore, the velocities can be expressed only in terms of the densities of quasimomenta at the corresponding Fermi quasimomenta; see Eq.~(\ref{eqn:BetheResc}). The same is true for the Drude weight matrix elements as Eq.~(\ref{eq:Dbfgen}) leads to
\begin{align}
\mathcal{D}_{\FB}=\frac{\pi\hbar n}{m} \left(1+\alpha-\alpha \frac{\varsigma(1)}{\varrho(1)}-\alpha \frac{\varrho(1)}{\varsigma(1)}\right).
\end{align}
The other matrix elements follow from the sum rules~(\ref{eq:sumrulesDrude}).

We have studied the exact decomposition of the Bethe-ansatz solution of the Bose--Fermi mixture for a special set of parameters defined by Eq.~(\ref{eq:condition}), in terms of the Lieb--Liniger model and the Yang--Gaudin model of attractive spin-$\frac{1}{2}$ fermions. This enabled us to obtain explicit analytical results for the mixture based on the known results for the two other models \footnote{Note that an interested reader can straightforwardly obtain analytically further terms in the series expansions of physical quantities once $e_{\L}$ and $e_{\Y}$ are used with more terms in their expansions.}. At weak interactions, the special set of parameters describes a system of free fermions and weakly interacting bosons that are weakly coupled. The fermion density is such that the Fermi velocity of the isolated system of fermions is equal to the sound velocity of the isolated system of weakly interacting bosons. However, due to the boson-fermion interaction, a splitting of the two velocities occurs. It is described by the subleading terms in Eq.~(\ref{eqn:eta1_v_wk}), which show that the velocity splitting scales in an unusual way, as $\gamma^{3/4}$. 

The latter result can be obtained using a simple bosonization approach \cite{cazalilla_instabilities_2003} that reduces to the statement that the squares of the velocities are the eigenvalues of the matrix
\begin{align}\label{eq:lastmatrix}
\frac{1}{m}\begin{pmatrix}
		n_{\F}\frac{\partial \mu_{\F}}{\partial n_{\F}} & 		n_{\F}\frac{\partial \mu_{\F}}{\partial n_{\B}} \\
		n_{\B}\frac{\partial \mu_{\B}}{\partial n_{\F}} & 		n_{\B}\frac{\partial \mu_{\B}}{\partial n_{\B}}
\end{pmatrix}.
\end{align}
The diagonal matrix elements in Eq.~(\ref{eq:lastmatrix}) correspond to the squares of the velocities of free fermions and weakly interacting bosons. On the other hand, the off-diagonal matrix elements account for the density-density interaction between the two subsystems. Using the ground-state energy in the mean-field approximation,
\begin{align}\label{eq:mfenergy}
\frac{E_0}{L}=\frac{\hbar^2}{2m}\left(\frac{\pi^2}{3} n_{\F}^3+ c n_{\B}^2 +2c n_{\F}n_{\B}\right), 
\end{align}
we find that the diagonal matrix elements in Eq.~(\ref{eq:lastmatrix}) are  $(\pi \hbar n_{\F}/m)^2$ and $(\hbar \sqrt{c n_{\B}}/m)^2$. Applying Eq.~(\ref{eq:condition}) imposes their equality, yielding $\alpha=1-\sqrt\gamma/\pi$ at $\gamma\ll 1$, cf.~Eq.~(\ref{eqn:gamma_wk_fn_alpha}). The diagonalization of the matrix (\ref{eq:lastmatrix}) then gives the result (\ref{eqn:eta1_v_wk}), where only the first correction should be retained. Note that the mean-field energy (\ref{eq:mfenergy}) differs from the exact result for the energy determined by Eq.~(\ref{eqn:eta1_en_wk_alpha}) in the subleading-order term. The corresponding difference term describes the first energy correction of weakly-interacting bosons; it is proportional to $\gamma^{3/2}$ \cite{lieb_exact_1963a}. The latter, however, does not participate in the velocity splitting (i.e., in its leading-order contribution). On the other hand, the coupling between the subsystems that is described by the last term in Eq.~(\ref{eq:mfenergy}), is on the same order of $\gamma^{3/2}$ and must be accounted for as it determines the velocity splitting.

At this point we must be aware that the squared velocities in the Bose--Fermi mixture (and likely more generally) are the eigenvalues of the product $\mathcal{D} \mathcal{M}$ \cite{soham_2026}, and not of the matrix (\ref{eq:lastmatrix}). Nevertheless the velocities obtained from Eq.~(\ref{eq:lastmatrix}) are correct at leading and subleading order. This originates from the smallness of the off-diagonal matrix element of $\mathcal{D}$, which is $\mathcal{D}_{\FB}\propto \gamma$;  see Eq.~(\ref{eqn:eta1_dff_wk}). Indeed, the Drude weight matrix can be decomposed as
\begin{align}\label{eq:DDDD}
\mathcal{D}=\frac{\pi\hbar}{m} \begin{pmatrix}
		n_{\F} & 		0 \\
		0 & 		n_{\B}
\end{pmatrix}+ \mathcal{D}_{\FB}\begin{pmatrix}
		- 1& 	1 \\
		1 & 		-1
\end{pmatrix}.
\end{align}
The product of the first matrix on the right-hand side of Eq.~(\ref{eq:DDDD}) and $\mathcal{M}$ gives Eq.~(\ref{eq:lastmatrix}). On the other hand, one can verify that the correction arising from the second term on the right-hand side of Eq.~(\ref{eq:DDDD}) modifies the velocities starting at next-to-subleading order. Thus the term proportional to $\mathcal{D}_{\FB}$ in Eq.~(\ref{eq:DDDD}) can be neglected in the present case of finding the velocity splitting. Such omission of $\mathcal{D}_{\FB}$, however, should be regarded as the exception rather than the rule \cite{chandak_unpub}.

\section*{Acknowledgments}
This project was supported in part by the Program
QuanTEdu-France ANR-22-CMAS-0001 France 2030.


%

\end{document}